\documentclass[prb,aps,twocolumn,amsmath,amssymb,floatfix,showpacs,
superscriptaddress]{revtex4}

\usepackage[dvips]{graphics}
\usepackage{color}
\definecolor{dred}{rgb}{0.6,0,0}

\begin{document}

\title{\textcolor{dred}{Spin-orbit interaction induced spin selective
transmission through a multi-terminal mesoscopic ring}}

\author{Moumita Dey}

\affiliation{Theoretical Condensed Matter Physics Division, Saha
Institute of Nuclear Physics, Sector-I, Block-AF, Bidhannagar,
Kolkata-700 064, India}

\author{Santanu K. Maiti}

\email{santanu.maiti@isical.ac.in}

\affiliation{Physics and Applied Mathematics Unit, Indian Statistical 
Institute, 203 Barrackpore Trunk Road, Kolkata-700 108, India}

\author{Sreekantha Sil}

\affiliation{Department of Physics, Visva-Bharati, Santiniketan, West 
Bengal-731 235, India}

\author{S. N. Karmakar}

\affiliation{Theoretical Condensed Matter Physics Division, Saha 
Institute of Nuclear Physics, Sector-I, Block-AF, Bidhannagar, 
Kolkata-700 064, India}

\begin{abstract}
Spin dependent transport in a multi-terminal mesoscopic ring is 
investigated in presence of Rashba and Dresselhaus spin-orbit 
interactions. Within a tight-binding framework we use a general spin 
density matrix formalism to evaluate all three components ($P_x$, $P_y$ 
and $P_z$) of the polarization vector associated with the charge current 
through the outgoing leads. It explores the dynamics of the spin 
polarization vector of current propagating through the system subjected 
to the Rashba and/or the Dresselhaus spin-orbit couplings. The sensitivity 
of the polarization components on the electrode-ring interface geometry 
is discussed in detail. Our present analysis provides an understanding
of the coupled spin and electron transport in mesoscopic bridge systems.
\end{abstract}

\pacs{72.25.-b, 73.23.-b, 85.35.Ds}

\maketitle

\section{Introduction}

One of the major goals of spintronic applications has always been to 
manipulate electron's spin degree of freedom to create a new 
paradigm~\cite{wolf} in the fields of quantum 
information processing. Spin-$\frac{1}{2}$ particles are a natural choice 
for a qubit in quantum computers. So, generation of spin polarized beam
is a highly significant issue as far as spintronic applications are 
concerned. A more or less usual way of realization~\cite{wang,xie} of 
spin filtering action is by using ferromagnetic leads or by external 
magnetic field. But, in the first case, spin injection from ferromagnetic 
lead is difficult due to large resistivity mismatch and for the second one, 
the difficulty is to confine a very strong magnetic field into a small 
region like a quantum ring. Therefore, attention is being paid for modeling 
of spin filter using the intrinsic properties of mesoscopic systems such as 
spin-orbit (SO) interaction~\cite{intrinsic1,intrinsic2,intrinsic3,
intrinsic4,intrinsic5,intrinsic6}. Originating from the relativistic 
correction to the Schr\"{o}dinger equation, SO interaction provides an all 
electrical way to generate and manipulate spin current in a far precise 
way rather than the usual magnetic field based spin control.

The main source of SO coupling in mesoscopic systems comes either from 
magnetic impurities (extrinsic type), or from bulk asymmetry or structural 
inversion asymmetry in the confining potential of the system (intrinsic 
type) yielding Dresselhaus or Rashba type of SO interaction~\cite{rashba,
dressel,winkler}. Studies on Rashba or Dresselhaus kind of interactions 
has made a significant impact in semiconductor spintronics as far as the 
control of spin dynamics is concerned.

Since SO interaction couples the spin degree of freedom with the momentum of 
an electron, so it is possible to achieve spin polarized currents in output 
terminals of a multi-terminal mesoscopic ring when an unpolarized electron 
beam is injected into its input terminal, though Kramer's degeneracy 
suggests that due to preservation of time reversal symmetry only SO 
interaction can never induce in a two-terminal system~\cite{kramer}, 
whereas, in case of multi-terminal system the condition gets relaxed. 
Additionally, in a multi-terminal system the degree of spin coherence can 
also be manipulated even when a polarized beam is injected.

Till date a lot of theoretical work has been done to model spin selective 
transmission. In 2003, Kislev and Kim have proposed that a planar T-shaped 
\begin{figure}[ht]
{\centering \resizebox*{7.5cm}{4cm}{\includegraphics{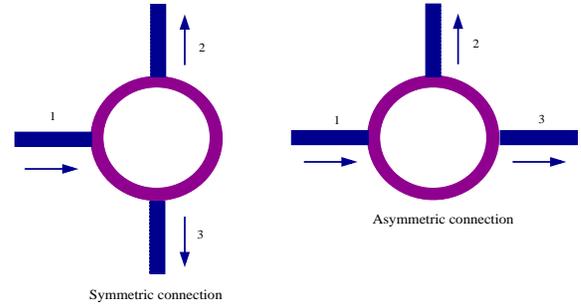}}\par}
\caption{(Color online). Schematic view of a mesoscopic ring, subjected
to Rashba and Dresselhaus SO interactions, with one input and two output 
terminals, where the arrows represent the movement of electrons. Two 
different configurations of electrode-ring interface geometry are taken
to explore quantum interference effect on spin polarization components.}
\label{multiring}
\end{figure}
structure with a ring resonator can be highly efficient in producing spin 
polarized currents in different output arms in presence of Rashba SO 
interaction~\cite{kim}. Following that in 2005, Shelykh {\em et al.} 
investigated 
analytically by S-matrix theory both the effects of magnetic flux and 
Rashba coupling on charge and spin transport in a multi-terminal quantum
ring~\cite{shelykh}. Then, in 2006, Peeters {\em et al.} have shown that a 
mesoscopic semiconductor ring with one input and two outputs can act as 
an electron spin beam splitter due to Quantum interference effect and 
Rashba SO interaction~\cite{peeters}. In another work, Rabani {\em et al.} 
has designed a spin filter and spin splitter considering a 
two-terminal device in presence of external magnetic field~\cite{hodd}.

Thus, the studies involving spin dependent transport in multi-terminal 
mesoscopic rings have already generated a wealth of literature, but to the 
best of our knowledge there is still a need to look deeper into the problem 
to address several important issues those have not been well analyzed 
earlier, for example, the understanding of the effect of Rashba and 
Dresselhaus SO couplings on all three components $P_x$, $P_y$ and $P_z$ 
of the polarization vector associated with the charge current through the 
outgoing leads, and also the sensitivity of the polarization components on 
different electrode-ring interface configurations.

In the present work we mainly concentrate on these issues. Here we
analyze the dependence of the polarization components $P_x$, $P_y$ and
$P_z$ in the two output terminals of a three-terminal mesoscopic ring
subjected to Rashba and Dresselhaus SO interactions. Within a
tight-binding (TB) framework we evaluate these quantities ($P_x$, $P_y$
and $P_z$) using a general spin density matrix formalism, which was first 
approached by Nikolic and co-workers~\cite{niko1} in 2004. They 
presented an elegant way of expressing all three components $P_x$,
$P_y$ and $P_z$ of outgoing spin polarization vector in terms of spin
resolved transmission matrices within the framework of Landauer-B\"{u}ttiker 
formalism. This formalism encompasses both pure and mixed states as incoming 
beam and provides knowledge to investigate how polarization 
evolves~\cite{niko2,niko3} in outgoing current due to Rashba or Dresselhaus 
interaction, scattering at lead-conductor interface and spin dependent 
scattering off impurities in both the weak and strong disordered regimes.
The sensitivity of the polarization components on the electrode-ring 
interface geometry is also described in detail to make the present 
communication a self contained study.

The rest of the paper is organized as follows. In Section II, we present 
the model and theoretical formulation to obtain spin polarization 
components of the output currents in terms of transmission probabilities. 
The numerical results are illustrated in Section III. Finally, in Section 
IV, the results are summarized.

\section{Theoretical framework}

In this section we describe the model quantum system within a TB framework 
and express the spin polarization components $P_x$, $P_y$ and $P_z$ in terms 
of transmission coefficients of outgoing electrons through the quantum ring
following spin density matrix formalism.

\subsection{Model and Hamiltonian}

Let us start with Fig.~\ref{multiring} where a mesoscopic ring subjected
to Rashba and Dresselhaus SO interactions is attached with one input and
two output terminals. A simple lattice model within the framework of 
TB approximation assuming only nearest-neighbor coupling is used to 
describe the ring and side-attached leads. The TB Hamiltonian describing 
the entire system gets the form:
\begin{equation}
H=H_{\mbox{\tiny ring}} + H_{\mbox{\tiny leads}} + H_{\mbox{\tiny tun}}.
\label{eqn1}
\end{equation}
The first term $H_{\mbox{\tiny ring}}$ describes the Hamiltonian of the 
ring and for a $N$-site ring it reads,
\begin{equation}
H_{\mbox{\tiny ring}} = \sum_{n=1}^{N} \mbox{\boldmath$c$}_{n}^{\dag} 
\mbox{\boldmath$\epsilon$} \mbox{\boldmath$c$}_{n} - \sum_{n=1}^{N}
\left( \mbox{\boldmath$c$}_{n}^{\dag} \mbox{\boldmath$t$}_{n,n+1}
\mbox{\boldmath$c$}_{n+1} + h.c.\right)
\label{eqn2}
\end{equation}
where, \\
$\mbox{\boldmath $c$}^{\dagger}_{n}=\left(\begin{array}{cc}
c_{n,\uparrow}^{\dagger} & c_{n,\downarrow}^{\dagger} 
\end{array}\right);$
$\mbox{\boldmath $c$}_{n}=\left(\begin{array}{c}
c_{n,\uparrow} \\
c_{n,\downarrow}\end{array}\right);$
and
$\mbox{\boldmath $\epsilon$}=\left(\begin{array}{cc}
\epsilon & 0 \\
0 & \epsilon \end{array}\right)$.\\
\vskip 0.2cm
\noindent
Here, $c_{n,\sigma}^{\dag}$ and $c_{n,\sigma}$ are the creation and 
annihilation operators, respectively, at the $n$-th site for an electron 
with spin $\sigma$ ($\uparrow$, $\downarrow$). $\epsilon$ being the on-site 
energy.

The factor $\mbox{\boldmath$t$}_{n,n+1}$ is the sum of three terms 
as follows.
\begin{equation}
\mbox{\boldmath$t$}_{n,n+1} = \mbox{\boldmath$t$}_{n,n+1}^{0} + 
\mbox{\boldmath$t$}_{n,n+1}^{R} + \mbox{\boldmath$t$}_{n,n+1}^{D}.
\label{eqn3}
\end{equation}
Here~\cite{san1,san2,san22,san3},\\
$\mbox{\boldmath $t$}_{n,n+1}^0=t \left(\begin{array}{cc}
1 & 0 \\
0 & 1 \end{array} \right)$, \\
$\mbox{\boldmath $t$}_{n,n+1}^R = -i t_R \left( 
\cos{\frac{\varphi_n + \varphi_{n+1}}{2}} \mbox{\boldmath$\sigma_x$} +
\sin{\frac{\varphi_n + \varphi_{n+1}}{2}} \mbox{\boldmath$\sigma_y$}
\right)$, \\
$\mbox{\boldmath $t$}_{n,n+1}^D = -i t_D \left( 
\sin{\frac{\varphi_n + \varphi_{n+1}}{2}} \mbox{\boldmath$\sigma_x$} +
\cos{\frac{\varphi_n + \varphi_{n+1}}{2}} \mbox{\boldmath$\sigma_y$}
\right)$.\\
\vskip 0.1cm
\noindent
In these above expressions $t$ is the isotropic nearest-neighbor coupling 
strength, whereas, 
$\mbox{\boldmath$t$}$$_{n,n+1}^{R}$ and $\mbox{\boldmath$t$}$$^D_{n,n+1}$ 
correspond to the spin 
dependent terms with $t_R$ and $t_D$ being the nearest-neighbor hopping 
integrals due to Rashba and Dresselhaus SO interactions, respectively, which 
introduce spin flipping in the system and $\varphi_n$ is the azimuthal angle
for the $n$-th site. Mathematically, it can be expressed as
$\varphi_n=2 \pi (n-1)/N$.

In our formulation we assume that the one-dimensional ($1$D) semi-infinite 
leads are free from any kind of disorder and SO interactions. They can be
expressed as,
\begin{equation}
H_{\mbox{\tiny leads}} = \sum_{\alpha} H_{\alpha} 
\label{eqn4}
\end{equation}
where,
\begin{equation}
H_{\alpha} = \sum \limits_{n}\epsilon_{l} c_{n}^{\dag}c_n  + 
\sum_{\langle mn \rangle} t_l c^{\dag}_{m} c_n.
\label{eqn5}
\end{equation}
Similarly, the ring-to-lead coupling is described by the following 
Hamiltonian.
\begin{equation}
H_{\mbox{\tiny tun}} = \sum_{\alpha} H_{{\mbox{\tiny tun}},\alpha}.
\label{eqn6}
\end{equation}
Here,
\begin{equation}
H_{{\mbox{\tiny tun}}, \alpha} = t_c[c^{\dag}_{i} c_m + c^{\dag}_{m} c_i].
\label{eqn7}
\end{equation}
In these above equations (Eqs.~\ref{eqn4}-\ref{eqn7}), the index 
$\alpha$ signifies the number of leads attached to the ring. It can be two
or three or even more depending on the number of outgoing leads in addition 
to the incoming one. $\epsilon_l$ and $t_l$ stand for the site energy and 
nearest-neighbor hopping between the sites of the leads. The coupling 
between the leads and the ring is denoted by the hopping integral $t_c$. 
In Eq.~\ref{eqn7}, $i$ and $m$ belong to the boundary sites of the ring 
and the leads, respectively.

\subsection{A brief introduction to spin density matrix formalism}

Most of the quantum interference phenomena observed in different experiments,
e.g., Aharonov-Bohm effect, weak localization effect, etc., within the 
mesoscopic regime deal with the aspect of orbital quantum coherence of
electronic states. At much low temperatures ($T<1\,$K) and for the systems
having $L < L_\phi$ ($L_\phi \simeq 1\mu$m), inelastic scattering processes 
get suppressed so that an electron can be described by a single orbital 
wave function within the system. Now if the spin degree of freedom of the
electron is taken into account then two separate vector spaces are multiplied 
tensorially to get the full Hilbert space of the quantum states i.e., 
\begin{equation}
\mathcal{H} = \mathcal{H}_0 \otimes \mathcal{H}_s.
\label{eqn8}
\end{equation}
Here, $\mathcal{H}_0$ spans over the orbital degrees of freedom while
$\mathcal{H}_s$ operates in spin space only. Therefore, any arbitrary state 
$|\psi \rangle \in \mathcal{H}$ can be written as a linear combination 
of $|\phi_{\alpha} \rangle \otimes |\sigma \rangle$, where 
$|\phi_{\alpha} \rangle$'s are the basis vectors of $\mathcal{H}_0$ and 
$|\sigma \rangle$'s are the eigenstates of $\vec{\sigma}.\hat{u}$.
$\vec{\sigma}$ being the Pauli spin matrix and $\hat{u}$ is the unit vector 
along the direction of spin quantization axis. Thus, we can write the most 
general form of $|\psi \rangle$ as,
\begin{equation}
|\psi \rangle = \sum_{\alpha, \sigma} C_{\alpha, \sigma} |\phi_{\alpha} 
\rangle \otimes |\sigma \rangle.
\label{eqn9}
\end{equation}
Here, we choose the quantization direction along the $+$ve $Z$ direction,
and accordingly, $|\sigma \rangle$'s are the eigenstates of 
$\sigma_z$ operator i.e., $| \uparrow \rangle$ $\equiv$ 
$\left(\begin{array}{c}
1 \\
0\end{array}\right)$ and $| \downarrow \rangle$ $\equiv$
$\left(\begin{array}{c}
0 \\
1\end{array}\right)$.

The corresponding density matrix operator~\cite{niko3,ballen} for the state 
$|\psi\rangle$ becomes,
\begin{equation}
\rho = |\psi \rangle \langle \psi|.
\label{eqn10}
\end{equation}
This state can describe a pure one or a mixture of different quantum
states. Below we consider both these two cases to get a complete picture. \\
\vskip 0.01cm
\noindent
$\bullet$ {\bf Coherent beam of electrons:} First, we consider the case 
where the state $|\psi \rangle$ is a pure one indicating a coherent beam
of electrons. 

If the electron is free from any kind of spin-dependent interactions, then 
spin and charge coherences are independent of each other, resulting a 
separable state $|\psi \rangle$. In that case $| \psi \rangle$ can be 
written as, 
\begin{equation}
| \psi \rangle = | \Phi \rangle \otimes | \Sigma \rangle
\label{eqn11}
\end{equation}
where, $|\Phi \rangle$ and $|\Sigma \rangle$ are the orbital and spin parts,
respectively, of the total wave function $|\psi \rangle$ and they are
expressed as follows.
\begin{eqnarray}
|\Phi \rangle & = & \sum_{\alpha} a_{\alpha} |\phi_{\alpha} \rangle, 
\nonumber\\
|\Sigma \rangle & = & (a_{\uparrow} |\uparrow \rangle + 
a_{\downarrow} |\uparrow \rangle).
\label{eqn12}
\end{eqnarray}
The corresponding density matrix operator is thus given by,
\begin{eqnarray}
\rho & = & |\psi \rangle \langle \psi| \nonumber\\
& = & |\Phi \rangle \langle \Phi| \otimes 
|\Sigma \rangle \langle \Sigma|\nonumber\\
& = & \rho_0 \otimes \rho_s.
\label{eqn13}
\end{eqnarray}
Here, $\rho_0$ and $\rho_s$ are the reduced density matrices in the orbital 
and spin spaces, respectively.

Now, in presence of SO interaction, the state $|\psi \rangle$ is no longer 
separable, and hence, individual orbital and spin parts lose their coherence 
and become entangled and the state $|\psi \rangle$ can be written 
in a form as expressed in Eq.~\ref{eqn9}.

In this situation,
\begin{eqnarray}
\rho & = & |\psi \rangle \langle \psi| \nonumber\\
& \neq & \rho_0 \otimes \rho_s.
\label{eqn14}
\end{eqnarray}
$\bullet$ {\bf Incoherent beam of electrons:} Next, we consider the case 
where the electronic beam is an incoherent mixture i.e., statistical 
superposition of different quantum states. Here, the state cannot be 
expressed like Eqs.~\ref{eqn9} and \ref{eqn11}.

In this situation the state is best represented by the density 
matrix operator as,
\begin{equation}
\rho = \sum_{i} w_i |\psi_i \rangle \langle \psi_i|
\label{eqn15}
\end{equation}
where, $w_i$ gives the probability for the ensemble to be found in the 
quantum state $|\psi_i \rangle$. 

In the present work, we want to determine the components $P_x$, $P_y$
and $P_z$ of output currents propagating through the leads attached to 
the ring subjected to SO interactions. Using Pauli spin operators they 
are expressed as follows.
\begin{equation}
P_x = \langle \sigma_x \rangle \nonumber \; ; 
P_y = \langle \sigma_y \rangle \nonumber \; ;
P_z = \langle \sigma_z \rangle. 
\end{equation}

Now, following quantum statistics, measurement of any spin observable 
$O_s$ is accomplished by the following way,
\begin{equation}
\langle O_s \rangle = Tr[\mbox{\boldmath$\rho$}_s 
\mbox{\boldmath$O$}_s]
\label{eqn16}
\end{equation}
where $\mbox{\boldmath$O$}$$_s$ is the matrix form of the operator $O_s$
ans $\mbox{\boldmath$\rho$}$$_s$ is the spin density matrix which is 
obtained by taking partial trace over the orbital degrees of freedom 
of the full density matrix $\mbox{\boldmath$\rho$}$,
\begin{eqnarray}
\mbox{\boldmath$\rho$}_s & = & Tr_o [\mbox{\boldmath$\rho$}] \nonumber\\
& = & \sum_{\alpha} \langle \phi_{\alpha}|\mbox{\boldmath$\rho$}|
\phi_{\alpha} \rangle.
\label{eqn17}
\end{eqnarray}
Thus, in our case spin density matrix plays the central role in 
understanding the quantum dynamics of a spin sub-system, subjected to 
SO interactions and attached to the environment through ideal (free from 
any kind of charge or spin dependent interaction) leads. For electrons
(i.e., spin-$1/2$ particles) $\rho_s$ has a simple $2 \times 2$ 
representation in a chosen basis $| \uparrow \rangle$, 
$| \downarrow \rangle$ $\in$ $\mathcal{H}_s$ which reads, 
\begin{equation}
\mbox{\boldmath$\rho$}_s = \left(\begin{array}{cc}
\rho_{\uparrow \uparrow} & \rho_{\uparrow \downarrow} \\
\rho_{\downarrow \uparrow} & \rho_{\downarrow \downarrow}\end{array}\right).
\label{eqn18}
\end{equation}
Here, the diagonal elements represent the probabilities of finding an 
electron with spin $| \uparrow \rangle$ or $| \downarrow \rangle$, whereas 
the off-diagonal elements describe the probabilities of coherent superposition 
of $| \uparrow \rangle$ and $| \downarrow \rangle$ states due to quantum 
interference effect. 

It can also be represented as,
\begin{equation}
\mbox{\boldmath$\rho$}_s = \frac{I_s + \vec{\mbox{\boldmath$P$}}.\vec{\mbox
{\boldmath$\sigma$}}}{2}.
\label{eqn19}
\end{equation}
Here $\vec{\textbf{P}}$ is the polarization vector whose components are 
evaluated from the following relations.
\begin{eqnarray}
P_x & = & Tr[\mbox{\boldmath$\rho$}_s \mbox{\boldmath$\sigma_x$}], \nonumber\\
P_y & = & Tr[\mbox{\boldmath$\rho$}_s \mbox{\boldmath$\sigma_y$}], \nonumber\\
P_z & = & Tr[\mbox{\boldmath$\rho$}_s \mbox{\boldmath$\sigma_z$}].
\label{eqn20}
\end{eqnarray}
For a completely unpolarized electron beam i.e., an incoherent mixture of 
up and down spin electrons, $|\vec{\textbf{P}}|$ = $0$. In this case, 
the spin density operator ($\rho_s$) becomes,
\begin{equation}
\rho_s = \rho_{\uparrow \uparrow} |\uparrow \rangle 
\langle \uparrow|
+ \rho_{\downarrow \downarrow} |\downarrow \rangle \langle \downarrow|.
\label{eqn21}
\end{equation}
It is seen that for an incoherent mixture of up and down spin electrons
the off-diagonal elements of $\rho_s$ are zero.

Therefore, to determine the polarization components of outgoing currents 
we need to construct the spin density matrix for the outgoing charge
current.

\subsection{General expressions of $P_x$, $P_y$ and $P_z$ in terms of 
transmission matrices for a mesoscopic ring coupled to source and drain 
leads having $M$ channels in each lead}

In this sub-section we present a general scheme for evaluating spin
polarization components of the charge current through the outgoing
leads, considering a mesoscopic ring subjected to SO interactions,
where each lead contains $M$ number of channels. 
 
In our problem, we assume that a beam of incoherent i.e., unpolarized 
electrons is injected from the source to the ring. Therefore, the 
incident beam is best represented by the spin density operator as,
\begin{equation}
\rho_{\mbox{\tiny{s}}}^{\mbox{\tiny{in}}} = 
n_{\uparrow} |\uparrow 
\rangle \langle \uparrow| 
+  n_{\downarrow} |\downarrow \rangle \langle \downarrow|.
\label{eqn22}
\end{equation}
For a fully unpolarized beam, $n_{\uparrow}$=$n_{\downarrow}$=$\frac{1}{2}$
and hence the polarization components are evaluated as-
(using En.~\ref{eqn20}):
\begin{eqnarray}
P_x & = & 0, \nonumber\\
P_y & = & 0, \nonumber\\
P_z & = & (n_{\uparrow} - n_{\downarrow}) = 0. \nonumber
\end{eqnarray}
Now, in presence of SO interaction, individual spin and orbital parts become
entangled, and thus, the corresponding outgoing beam through the lead j,
can be described in terms of reduced spin density matrix~\cite{niko2} by 
taking partial trace over all the orbital degrees of freedom.
\begin{equation}
\mbox{\boldmath$\rho$}^{\mbox{\tiny{out}}}_{\mbox{\tiny{s,j}}} = 
\frac{e^2/h}{\zeta} 
\sum_{n^{\prime},n=1}^M 
\left(\begin{array}{cc}
\alpha & \beta \\
\gamma &
\delta
\end{array}\right)
\label{eqn23}
\end{equation}
where, 
\vskip 0.1cm
\noindent
$\zeta=n_{\uparrow} (G^{\uparrow \uparrow}_{ji} + G^{\downarrow 
\uparrow}_{ji} ) + n_{\downarrow} (G^{\uparrow \downarrow}_{ji} + 
G^{\downarrow \downarrow}_{ji})$, 
\vskip 0.1cm
\noindent
$\alpha=n_{\uparrow} |[\mbox{\boldmath$t^{\uparrow\uparrow}_{ij}$}]_{n'n}|^2 
+ n_{\downarrow} |[\mbox{\boldmath$t_{ji}^{\uparrow \downarrow}$}]_{n'n}|^2$,
\vskip 0.1cm
\noindent
$\beta=n_{\uparrow} [\mbox{\boldmath$t_{ji}^{\uparrow \uparrow}$}]_{n'n} 
[\mbox{\boldmath$t_{ji}^{\downarrow \uparrow}$}]_{n'n}^{*} +
n_{\downarrow} [\mbox{\boldmath$t_{ji}^{\uparrow \downarrow}$}]_{n'n} 
[\mbox{\boldmath$t_{ji}^{\downarrow \downarrow}$}]_{n'n}^{*}$, 
\vskip 0.1cm
\noindent
$\gamma=n_{\uparrow} [\mbox{\boldmath$t_{ji}^{\uparrow \uparrow}$}]_{n'n}^{*} 
[\mbox{\boldmath$t_{ji}^{\downarrow \uparrow}$}]_{n'n} +
n_{\downarrow} [\mbox{\boldmath$t_{ji}^{\uparrow \downarrow}$}]_{n'n}^{*} 
[\mbox{\boldmath$t_{ji}^{\downarrow \downarrow}$}]_{n'n}$,
\vskip 0.1cm
\noindent
$\delta=n_{\uparrow} |[\mbox{\boldmath$t_{ji}^{\downarrow 
\uparrow}$}]_{n'n}|^2 + n_{\downarrow} |[\mbox{\boldmath$t_{ji}^{\downarrow 
\downarrow}$}]_{n'n}|^2$.
\vskip 0.3cm
\noindent
Following the prescription of En.~(\ref{eqn20}), the spin polarization 
components of the outgoing through lead j, current can be expressed as, 
\begin{eqnarray}
P_j^{x} & = & \frac{2 e^2/h}{G_{ji}} 
\mbox{Re}\left[ \mbox{Tr} \left[[\mbox{\boldmath$t^{\uparrow \uparrow}_{ji}$}]
[\mbox{\boldmath$t^{\downarrow \uparrow}_{ji}$}]^{\dag} + 
[\mbox{\boldmath$t^{\uparrow \downarrow}_{ji}$}]
[\mbox{\boldmath$t^{\downarrow \downarrow}_{ji}$}]^{\dag}
\right]\right], \nonumber\\
P_j^{y} & = &  \frac{2 e^2/h}{G_{ji}} 
\mbox{Im}\left[\mbox{Tr}\left[[\mbox{\boldmath$t^{\uparrow 
\uparrow}_{ji}$}]^{\dag}
[\mbox{\boldmath$t^{\downarrow \uparrow}_{ji}$}] + 
[\mbox{\boldmath$t^{\uparrow \downarrow}_{ji}$}]^{\dag}
[\mbox{\boldmath$t^{\downarrow \downarrow}_{ji}$}]\right]\right], \nonumber\\
P_j^{z} & = & \frac{G^{\uparrow \uparrow}_{ji} + G^{\uparrow \downarrow}_{ji} - 
G^{\downarrow \uparrow}_{ji} - G^{\downarrow \downarrow}_{ji}}
{G^{\uparrow \uparrow}_{ji} + G^{\uparrow \downarrow}_{ji} + 
G^{\downarrow \uparrow}_{ji} + G^{\downarrow \downarrow}_{ji}}.
\label{eqn24}
\end{eqnarray}
where, $G_{ji}$=$(G_{ji}^{\uparrow \uparrow}+G_{ji}^{\uparrow \downarrow}
+G_{ji}^{\downarrow \uparrow}+G_{ji}^{\downarrow \downarrow})$.

Here, $[\mbox{\boldmath$t_{ji}^{\sigma' \sigma}$}]$ is the transmission 
matrix, having dimension $M \times M$, for an electron injected from the lead 
i, with spin $\sigma$ and transmitted through the lead j with $\sigma^{'}$. 
Thus, for a single channel lead $[\mbox{\boldmath$t_{ji}^{\sigma' \sigma}$}]$ 
becomes a simple element rather than being a matrix. 
$G_{ji}^{\sigma' \sigma}$ is the conductance of the ring, defined by 
the Landauer formula~\cite{land}, and at low bias voltage it gets the form,
\begin{equation}
G_{ji}^{\sigma' \sigma} = \frac{e^2}{h}
\mbox{Tr}\left[ \mbox{\boldmath$[t_{ji}^{\sigma' \sigma}]
[t_{ji}^{\sigma' \sigma}$}]^{\dag}\right].
\label{eqn25}
\end{equation}
Below we describe the way of determining the transmission matrix and its 
relation to the conductance.

\subsection{Evaluation of transmission matrix}

The transmission matrix is obtained from the well known 
relation~\cite{lee,butt1,butt2,butt3},
\begin{equation}
\mbox{\boldmath$t_{ji}^{\sigma' \sigma}$} = \sqrt{\mbox
{\boldmath$\Gamma_{i,\mbox{\tiny{red}}}^{\sigma}$}} 
\mbox{\boldmath$\mathcal{G}_{ij}^{\sigma \sigma'}$} 
\sqrt{\mbox{\boldmath$\Gamma_{j,\mbox{\tiny{red}}}^{\sigma'}$}}.
\label{eqn26}
\end{equation}
Here, $\mbox{\boldmath$\mathcal{G}_{ij}^{\sigma \sigma'}$}$ is the retarded
Green's function (in matrix form) in the reduced dimension $M \times M$, 
connecting $i$-th and $j$-th leads (each lead contains $M$ channels) i.e., 
$\mbox{\boldmath$\mathcal{G}_{ij}^{\sigma \sigma'}$}$ = $\langle 
\textbf{i},\sigma| \mbox{\boldmath$\mathcal{G}$}| \textbf{j}, 
\sigma^{\prime} \rangle$. 
$\mbox{\boldmath$\Gamma$}_{i(j),\mbox{\tiny{red}}}^{\sigma (\sigma')}$'s are 
the coupling matrices in the same reduced dimension.

The single particle Green's function describing the complete system 
i.e., the ring with side-attached leads for an electron with energy 
$E$ is defined as,
\begin{equation}
G=\left(E-H + i\eta \right)^{-1}
\label{eqn27}
\end{equation}
where, $\eta \rightarrow 0^+$.

The problem of finding \mbox{\boldmath $G$} in the full Hilbert space of 
\mbox{\boldmath $H$} taking the matrix forms of \mbox{\boldmath $H$} and 
\mbox{\boldmath $G$} can be mapped exactly to a Green's function
\mbox{\boldmath $G_{\mbox{\tiny{ring}}}^{\mbox{\tiny{eff}}}$} which
represents an effective Hamiltonian in the reduced Hilbert space of the 
mesoscopic ring. It is expressed like,
\begin{equation}
\mbox{\boldmath ${\mathcal G}$=$G_{\mbox{\tiny{ring}}}^{\mbox{\tiny{eff}}}$}
=\left(\mbox{\boldmath $E- H_{\mbox{\tiny{ring}}}-\sum 
\limits_{\alpha, \sigma} \Sigma_{\alpha}^{\sigma}$}\right)^{-1}
\label{eqn28}
\end{equation}
where,
\begin{equation}
\mbox{\boldmath $\Sigma_{\alpha}^{\sigma}$} = \mbox{\boldmath 
$H_{\mbox{\tiny{tun}}, \alpha}^{\dag} G_{\alpha} 
H_{\mbox{\tiny{tun}}, \alpha}$}. 
\label{eqn29}
\end{equation}
These \mbox{\boldmath $\Sigma_{\alpha}^{\sigma}$}'s are the self-energies 
introduced to incorporate the effect of coupling of the ring to the leads. 
It is evident from Eq.~\ref{eqn29} that the form of the self-energies are 
independent of the ring itself through which spin transmission is studied. 
\mbox{\boldmath$\Gamma_{\alpha}^{\sigma}$}'s describe the coupling between 
the ring and the leads and they are mathematically defined as,
\begin{equation}
\mbox {\boldmath $\Gamma_{\alpha}^{\sigma}$} = i \left[\mbox 
{\boldmath $\Sigma_{\alpha}^{\sigma}-\Sigma_{\alpha}^{\sigma \dag}$}\right],
\label{eqn30}
\end{equation}
where, \mbox{\boldmath $\Sigma_{\alpha}^{\sigma}$} and 
\mbox{\boldmath $\Sigma_{\alpha}^{\sigma \dag}$}
are the retarded and advanced self-energies associated with the $\alpha$-th
lead, respectively. This self-energy term is again 
expressed~\cite{datta1,datta2} as a sum of real and imaginary parts,
\begin{equation}
\mbox{\boldmath ${\Sigma^{\sigma}_{\alpha}}$} = \mbox{\boldmath 
$\Lambda_{\alpha}^{\sigma}$} - i \mbox{\boldmath $\Delta_{\alpha}^{\sigma}$},
\label{eqn31}
\end{equation}
where, they describe the energy shift and broadening of the energy levels
of the ring, respectively. The finite imaginary part is obtained due to 
inclusion of semi-infinite leads having continuous energy spectrum. 
Therefore, the coupling matrices can easily be determined from the 
self-energy expression and is expressed in the form,
\begin{equation}
\mbox{\boldmath $\Gamma_{\alpha}^{\sigma}$}=-2\,{\mbox {Im}} 
(\mbox{\boldmath $\Sigma_{\alpha}^{\sigma}$}).
\label{eqn32}
\end{equation}
From this relation (Eq.~\ref{eqn32}) the reduced coupling matrices
\mbox{\boldmath$\Gamma_{\alpha,\mbox{\tiny{red}}}^{\sigma}$}'s are 
constructed.

Following the reference~\cite{niko4}, the self-energy matrices 
(\mbox{\boldmath$\Sigma^{\sigma}_{\alpha}$}'s) are evaluated in the reduced 
Hilbert space of the ring. The details are also available in other 
articles~\cite{mou1,mou2,other1,other2}.

Although the theoretical formulation presented above is based on a 
general approach considering finite width leads having $M$ number of
channels, for completeness of the theoretical description, but in this 
article we present all the numerical results (Sec. III) 
considering single-channel leads. 

\section{Numerical results and discussion}

In what follows we restrict ourselves to absolute zero temperature and use
the units where $e=h=c=1$. Throughout the numerical calculations we choose 
$\epsilon$ = $\epsilon_l$ = $0$ and $t=t_l=t_c=-1$. The energy scale is 
measured in unit of $t$ and the SO coupling strengths ($t_R$ and $t_D$) 
are also scaled in unit of $t$.

\subsection{Two-terminal transport}

Before addressing the central problem i.e., the possibilities of getting 
spin polarized currents in two output terminals of a three-terminal 
\begin{figure}[ht]
{\centering \resizebox*{7cm}{3.5cm}{\includegraphics{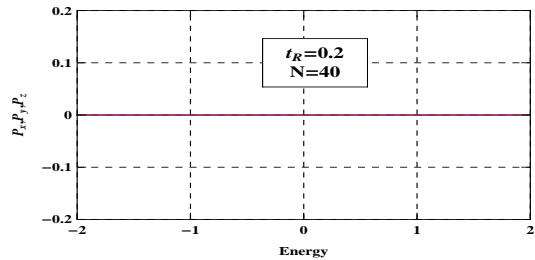}}\par}
\caption{(Color online). $P_x$, $P_y$ and $P_z$ as function of energy $E$
for a symmetrically connected mesoscopic ring with $N=40$, where we set
$t_R=0.2$ and $t_D=0$.}
\label{poltwosym1}
\end{figure}
mesoscopic ring from a completely unpolarized incident electron beam, 
first we analyze the results for a simple system where a mesoscopic ring, 
subjected to Rashba and Dresselhaus SO interactions, is coupled to two leads.
In Fig.~\ref{poltwosym1} we show the variation of spin polarization 
components $P_x$, $P_y$ and $P_z$ as a function of energy $E$ for a
symmetrically connected two-terminal mesoscopic ring in presence of
a non-zero value of Rashba SO coupling strength. Here we take a 
$40$-site ring and set $t_R=0.2$. The Dresselhaus SO coupling is
fixed at zero. From this spectrum it is clearly observed that all
three polarization components of the outgoing current become exactly
zero for the entire energy band region. We also carry out extensive
numerical work considering different values of Rashba SO coupling (the 
results are shown in Fig.~\ref{poltwosym2}) and for other possible 
lead-ring interface geometries i.e., for asymmetrical connections and 
find the identical behavior of these three components. Thus we can 
emphasize that in absence of any magnetic impurity or external magnetic 
\begin{figure}[ht]
{\centering \resizebox*{7cm}{3.5cm}{\includegraphics{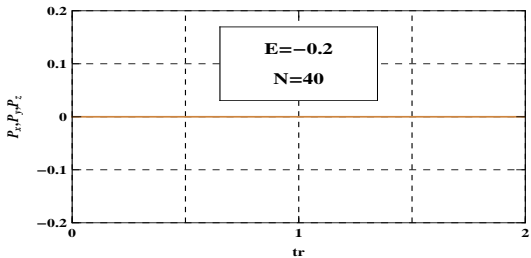}}\par}
\caption{(Color online). $P_x$, $P_y$ and $P_z$ as a function of Rashba SO
coupling strength $t_R$ at a particular energy $E=-0.2$ for the identical
lead-ring configuration taken in Fig.~\ref{poltwosym1}. The ring size $N$
and the Dresselhaus SO coupling strength are also same as 
Fig.~\ref{poltwosym1}.}
\label{poltwosym2}
\end{figure}
\begin{figure}[ht]
{\centering \resizebox*{7cm}{8cm}{\includegraphics{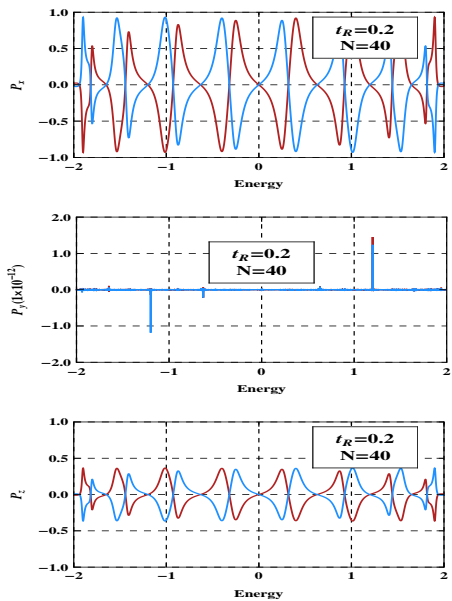}}\par}
\caption{(Color online). Polarization components of outgoing currents
as a function of energy in a three-terminal mesoscopic ring ($N=40$) 
when the output leads are connected symmetrically with respect to the 
source lead. The red and blue lines correspond to the results for the 
leads $2$ and $3$, respectively. Here we set $t_R=0.2$ and $t_D=0$.} 
\label{poleng1}
\end{figure}
field, only SO coupling cannot induce spin polarization in a two-terminal 
quantum system and this feature is independent of the lead-ring interface 
geometry. The reason is that in presence of SO interaction the 
time-reversal symmetry is not broken, and accordingly, the Kramer's 
degeneracy between the states 
$|k \uparrow \rangle$ and $|- k \downarrow \rangle$ gets preserved.
In a work Kim {\em et al.}~\cite{kim} have argued on the basis of 
symmetry of the S-matrix elements that a two-terminal time-reversal 
invariant system is incapable of producing spontaneous spin polarization. 
Moore {\em et al.}~\cite{kramer} have also shown in their recent work that 
only SO interaction cannot remove the degeneracies of the transmission 
eigenvalues. Thus, our numerical results exactly corroborate with these 
\begin{figure}[ht]
{\centering \resizebox*{7cm}{8cm}{\includegraphics{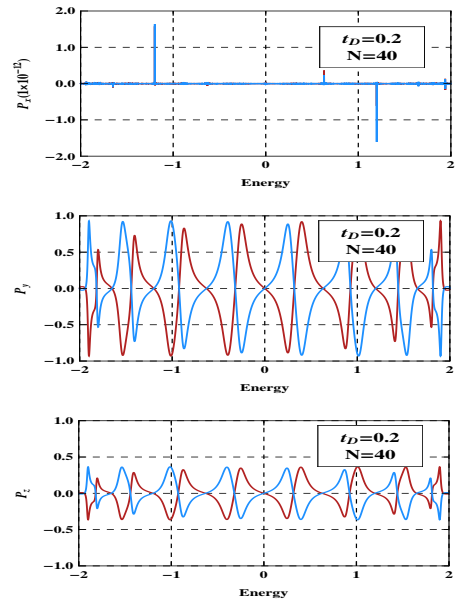}}\par}
\caption{(Color online). Same as Fig.~\ref{poleng1}, with $t_R=0$ and 
$t_D=0.2$.} 
\label{poleng2}
\end{figure}
arguments. A similar nature of the polarization components in this 
two-terminal geometry is also be obtained when they are plotted as a 
function of Dresselhaus SO coupling setting $t_R=0$. 

\subsection{Three-terminal transport}

In this sub-section we discuss the central results of our present 
investigation i.e., the interplay of Rashba and Dresselhaus SO couplings
\begin{figure}[ht]
{\centering \resizebox*{3.2cm}{3.5cm}{\includegraphics{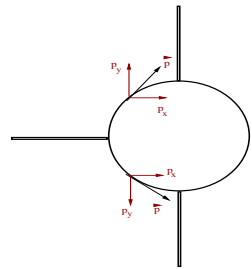}}\par}
\caption{(Color online). Momentum vectors and their components at the
two equivalent sites for the two arms of the ring.}
\label{diagram}
\end{figure}
and lead-ring interface geometry on the polarization components of 
outgoing currents in a three-terminal mesoscopic ring. We analyze the 
results for two distinct configurations of lead-ring geometry as
schematically illustrated in Fig.~\ref{multiring}. In one configuration
the outgoing leads are coupled symmetrically with respect to the source 
lead, while in the other case they are connected asymmetrically.

\subsubsection{Symmetric configuration}

We start by discussing the variation of spin polarization components 
$P_x$, $P_y$ and $P_z$ of outgoing currents in a three-terminal mesoscopic
\begin{figure}[ht]
{\centering \resizebox*{7cm}{6cm}{\includegraphics{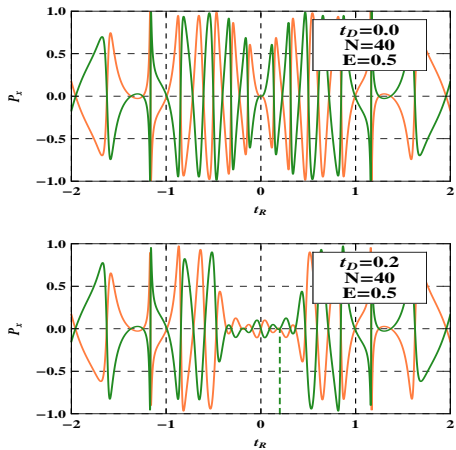}}\par}
\caption{(Color online). $P_x$ as a function of Rashba SO coupling in a 
three-terminal mesoscopic ring ($N=40$) for a typical energy $E=0.5$
when the output leads are attached symmetrically with respect to the 
source lead. The results are shown for two different values of $t_D$,
where the red and green curves represent the results for the leads $2$
and $3$, respectively.}
\label{polxr}
\end{figure}
\begin{figure}[ht]
{\centering \resizebox*{7cm}{6cm}{\includegraphics{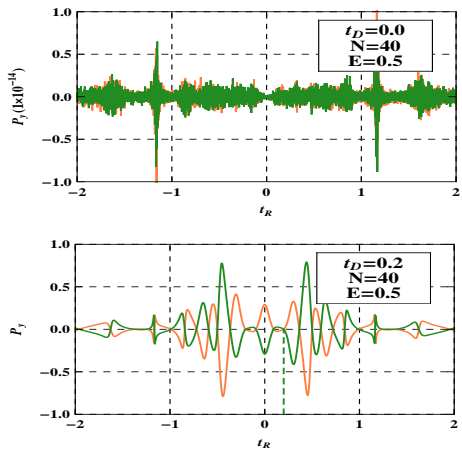}}\par}
\caption{(Color online). $P_y$ as a function of Rashba SO coupling.
The lead-ring interface geometry and all the other parameters are 
same as Fig.~\ref{polxr}.}
\label{polyr}
\end{figure}
ring with Rashba SO coupling only, that is, setting the Dresselhaus SO 
coupling to zero. The results for a $40$-site ring with $t_R=0.2$ and
$t_D=0$ are shown in Fig.~\ref{poleng1}, where the red lines describe 
the results for the output lead $2$, while for the other lead (lead $3$) 
they are presented by the blue lines. From the spectra it is observed 
that the $X$ and $Z$ components of the spin polarization vectors in two 
symmetrically coupled output leads are exactly identical in magnitude
but they carry opposite signs for each value of the injecting electron
energy $E$. On the other hand, the component $P_y$ exhibits identical
sign in both these two leads providing vanishingly small amplitudes.
These features can be explained from the following arguments. 

The $Z$ component of spin polarization essentially describes the normalized 
difference between the charge currents of up and down spin electrons 
\begin{figure}[ht]
{\centering \resizebox*{7cm}{6cm}{\includegraphics{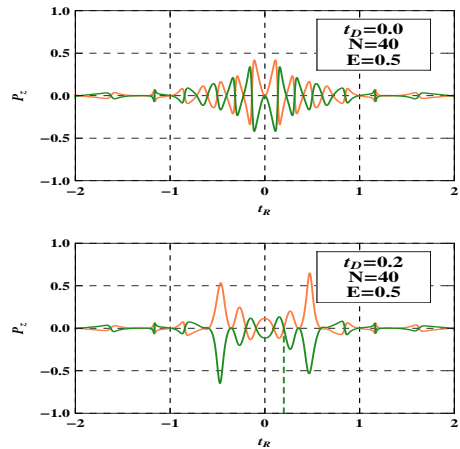}}\par}
\caption{(Color online). $P_z$ as a function of Rashba SO coupling.
The lead-ring interface geometry and all the other parameters are 
same as Fig.~\ref{polxr}.}
\label{polzr}
\end{figure}
flowing through the output leads (see Eq.~\ref{eqn24}), since in our
present scheme we choose the quantization direction along the $+$ve
$Z$ axis. Now, it is well known that the up and down spin electrons 
scatter in opposite directions when they traverse through a conductor
subjected to a SO interaction which is the aspect of visualizing mesoscopic
spin Hall effect and accumulation of opposite spins on the opposite 
edges. Therefore, spin polarization with opposite signs for the component
$P_z$ is expected in two output leads those are attached symmetrically to 
the mesoscopic ring with respect to the input lead. 

The above argument cannot be given to explain the characteristic features 
of the other two components $P_x$ and $P_y$ since we select $+$ve $Z$ axis 
as the quantization direction. In our theoretical framework we have already
stated that $P_x$ and $P_y$ are evaluated from the expectation
values of $\mbox{\boldmath$\sigma$}_x$ and $\mbox{\boldmath$\sigma$}_y$. 
Now, we know from the Rashba Hamiltonian that $\mbox{\boldmath$\sigma$}_x$ 
is coupled with the $Y$-component of the momentum of an electron which gets 
opposite signs at two equivalent atomic sites in the two arms of the ring 
resulting opposite spin 
polarization $P_x$ in the output leads. But, for $P_y$ the situation is
somewhat different as $\mbox{\boldmath$\sigma$}_y$ is coupled with the
$X$-component of the momentum which shows identical sign at 
the equivalent points (see fig.\ref{diagram}). Hence, a destructive 
interference takes place 
and it leads to a vanishingly small spin polarization at the output 
leads. Based on the symmetry arguments of the S-matrix elements 
Kim {\em et al.}~\cite{kim} have shown analytically that in a Y-shaped
conductor subjected to SO interaction the transmission amplitudes for the
$X$ and $Z$ components get equal magnitude and opposite phases for 
symmetrically connected leads which provide opposite spin polarization 
for these two components. While, for the $Y$ component almost zero 
polarization is achieved with identical phase in the leads. Our numerical
results for a three-terminal mesoscopic ring match very well with their 
analytical findings. 

Next, we focus our attention on the behavior of spin polarization components
of outgoing currents considering a mesoscopic ring with only Dresselhaus SO 
\begin{figure}[ht]
{\centering \resizebox*{7cm}{9cm}{\includegraphics{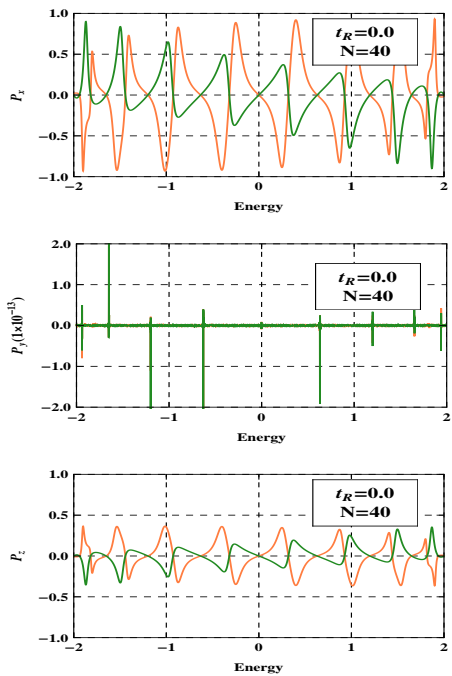}}\par}
\caption{(Color online). Polarization components of outgoing currents as
a function of energy in three-terminal mesoscopic ring ($N=40$) when the
output leads are coupled asymmetrically with respect to the source lead.
The red and green curves correspond to the results for the leads $2$ and
$3$, respectively. Here we choose $t_R=0.2$ and $t_D=0$.}
\label{polasymeng1}
\end{figure}
interaction. The results are presented in Fig.~\ref{poleng2}, where 
$P_x$, $P_y$ and $P_z$ are computed for the same ring size ($N=40$) and 
identical lead-ring interface geometry used in Fig.~\ref{poleng1}, setting
$t_R=0$ and $t_D=0.2$. Interestingly, we see that the $Z$ component
alternates its sign in two output leads keeping the magnitude unchanged,
while the other two components $P_x$ and $P_y$ interchange their features
compared to the previous case i.e., where the ring is described with Rashba
SO interaction only. Below we justify these phenomena through simple 
analytical arguments.

The TB Hamiltonians ($\mbox{\boldmath $H_R$}$ and $\mbox{\boldmath $H_D$}$) 
describing the Rashba and Dresselhaus SO interaction terms can be 
transformed into each other by a simple unitary transformation using the 
matrix $\mbox{\boldmath$U$}$, i.e.,  
$\mbox{\boldmath $U^{\dag} H_R U$}$ = $\mbox{\boldmath $H_D$}$, where
$\mbox{\boldmath $U$}$ = $\left(\mbox{\boldmath$\sigma$}_x 
+ \mbox{\boldmath$\sigma$}_y \right)/\sqrt{2}$. Therefore, if $|\psi\rangle$
is the eigenstate of $\mbox{\boldmath $H_R$}$ corresponding to a particular
energy eigenvalue and $|\psi^{\prime}\rangle$ is the eigenstate of the
transformed Hamiltonian $\mbox{\boldmath $H_D$}$, then we can write
$|\psi^{\prime} \rangle$ = $\mbox{\boldmath $U$} |\psi \rangle$. Following 
this transformation the Z component of spin polarization (Eq.~\ref{eqn20}) 
in presence of only Dresselhaus SO interaction gets the form: 
\begin{eqnarray}
P_z|_D & = & \langle\mbox{\boldmath$\sigma$}_z^{\prime}\rangle \nonumber\\
& = & \langle\psi^{\prime}|\mbox{\boldmath $\sigma$}_z|\psi^{\prime}\rangle 
\nonumber \\
& = & \langle\psi|\mbox{\boldmath $U$}^\dag \mbox{\boldmath$\sigma$}_z 
\mbox{\boldmath $U$}|\psi\rangle \nonumber \\
& = & \langle\psi|(-\mbox{\boldmath$\sigma$}_z)|\psi\rangle \nonumber \\
& = & -\langle\psi|\mbox{\boldmath$\sigma$}_z|\psi\rangle \nonumber \\
& = & -\langle\mbox{\boldmath$\sigma$}_z\rangle|\nonumber \\
& = & -P_z|_R.
\label{eqn33}
\end{eqnarray}
This equation (Eq.~\ref{eqn33}) clearly illustrates the reason behind the 
sign reversal of $P_z$ of outgoing currents in two leads when the ring
is described with only Dresselhaus SO term compared to the other case 
i.e., the ring with only Rashba SO interaction.   

Following the above prescription we can also get the relations
\begin{equation}
P_x|_D = P_y|_R~~~\mbox{and}~~~ P_y|_D = P_x|_R 
\label{eqn34}
\end{equation}
since,
\begin{eqnarray}
\mbox{\boldmath $U$}^{\dag}\mbox{\boldmath $\sigma$}_x\mbox{\boldmath $U$} 
=\mbox{\boldmath $\sigma$}_y~~~\mbox{and}~~~
\mbox{\boldmath$U$}^{\dag}\mbox{\boldmath$\sigma$}_y\mbox{\boldmath$U$} 
= \mbox{\boldmath$\sigma$}_x.
\label{eqn35}
\end{eqnarray}
These expressions (Eqs.~\ref{eqn34} and \ref{eqn35}) yield the reason for
interchanging the features of $P_x$ and $P_y$ in the mesoscopic ring with
only Dresselhaus SO coupling.

Finally, in Figs.~\ref{polxr}-\ref{polzr} we present the variations of
$P_x$, $P_y$ and $P_z$ as a function of Rashba SO coupling for a typical
energy $E=0.5$ considering the identical ring size used in Figs.~\ref{poleng1}
and \ref{poleng2}. The results are computed for two different values of 
$t_D$ to explore the combined effect of Rashba and Dresselhaus SO interactions
on the spin polarization components. When $t_D=0$, the components $P_x$ and
$P_z$ provide outgoing currents with equal magnitude and opposite phases,
while the other component $P_y$ almost drops to zero. The situation is 
somewhat interesting when Dresselhaus SO coupling is included in addition 
to the Rashba term. For the non-zero value of $t_D$, all these three
components get finite values in output leads and the magnitudes of 
individual components also differ in these two leads. These phenomena can 
be well explained from the analysis described above. Very interestingly
we see that, at the particular case when the strengths of these two SO
couplings are identical i.e., $t_R=t_D$, the three polarization components 
drop to zero. (shown by the green dashed lines in
Figs.~\ref{polxr}-\ref{polzr}).
It is already said that $P_z$ in presence of Rashba coupling is
just equal in magnitude but opposite in sign with $P_z$ in presence of 
Dresselhaus coupling, so when both the interactions are present in equal 
strength net polarization must vanish. For the other components we argue that 
when $t_R$ = $t_D$, the total Hamiltonian commutes
with $(\mbox{\boldmath$\sigma$}_x + \mbox{\boldmath$\sigma$}_y)$, so 
($\langle \sigma_x \rangle$ + $\langle \sigma_y \rangle$) i.e., 
($P_x$+$P_y$) is a conserved quantity, i.e., ($P_x$+$P_y$) should
have the same sign and same magnitude in both leads, which is possible
only when both $P_x$ and $P_y$ are individually zero at both the outputs.
\begin{figure}[ht]
{\centering \resizebox*{7cm}{9cm}{\includegraphics{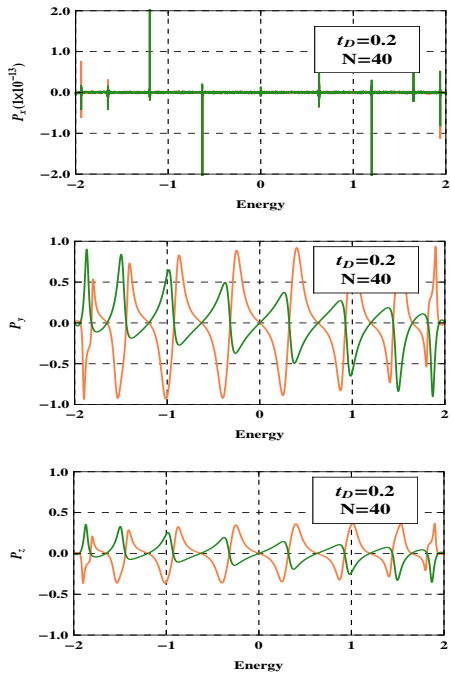}}\par}
\caption{(Color online). Same as Fig.~\ref{polasymeng1}, with $t_R=0$ and
$t_D=0.2$.}
\label{polasymeng2}
\end{figure}
This phenomenon emphasizes that any one
of the SO fields can be predicted precisely if the other one is known.
Needless to say, the precise determination of the SO coupling strengths
is extremely crucial in the field of spintronics. One can control Rashba
SO interaction by a suitable gate voltage, and hence, it can be measured. 
On the other hand, the possible techniques for the determination of 
Dresselhaus SO coupling are relatively few. Recently, we have put forward 
some ideas of estimating SO coupling strengths in a single sample which 
include the measurement of the minimum in Drude weight~\cite{san1} which
describes the conducting nature of the material, the observation of 
transmission resonance or anti-resonance of outgoing electrons~\cite{san2} 
and the determination of persistent spin current~\cite{san3} in an isolated 
loop geometry in presence of these two SO interactions. In the present work 
we give a separate proposal for it. For a single sample if a finite 
polarization is achieved for a particular value of $t_D$, then one can tune 
Rashba SO coupling by means of gate voltage to get vanishing spin 
polarization at the output currents. Thus, knowing the Rashba SO coupling, 
an accurate measurement of Dresselhaus term is possible.

\subsubsection{Asymmetric configuration}

The spin polarization turns out to be sensitive to the lead-ring 
\begin{figure}[ht]
{\centering \resizebox*{7cm}{6cm}{\includegraphics{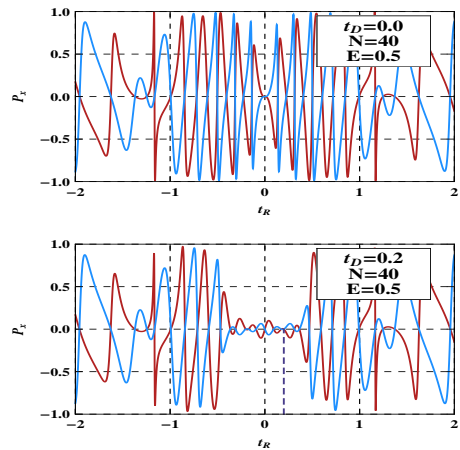}}\par}
\caption{(Color online). $P_x$ as a function of Rashba SO coupling in a 
three-terminal mesoscopic ring ($N=40$) for a typical energy $E=0.5$
when the output leads are attached asymmetrically with respect to the 
source lead. The results are shown for two different values of $t_D$,
where the red and blue curves represent the results for the leads $2$
and $3$, respectively.}
\label{polx}
\end{figure}
\begin{figure}[ht]
{\centering \resizebox*{7cm}{6cm}{\includegraphics{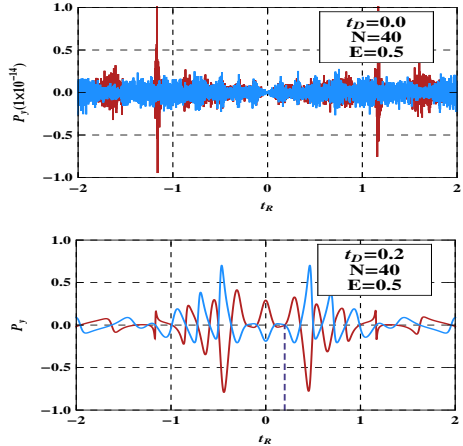}}\par}
\caption{(Color online). $P_y$ as a function of Rashba SO coupling. 
The lead-ring interface geometry and all the other parameters are 
same as Fig.~\ref{polx}.}
\label{poly}
\end{figure}
interface geometry. To this end, we analyze the behavior of spin 
polarization components in a three-terminal mesoscopic ring where the
output leads are coupled asymmetrically (see Fig.~\ref{multiring}) with 
respect to the source lead.

In Fig.~\ref{polasymeng1} we display the variation of spin polarization
components $P_x$, $P_y$ and $P_z$ for a $40$-site ring described with 
Rashba SO interaction only, i.e., setting $t_D$ at zero. The red curves
represent the results for the lead $2$, while for the lead $3$ they are
shown by the green lines. From the spectra we see that, unlike the 
symmetric configuration, the magnitudes of $P_x$ and $P_z$ in two output 
\begin{figure}[ht]
{\centering \resizebox*{7cm}{6cm}{\includegraphics{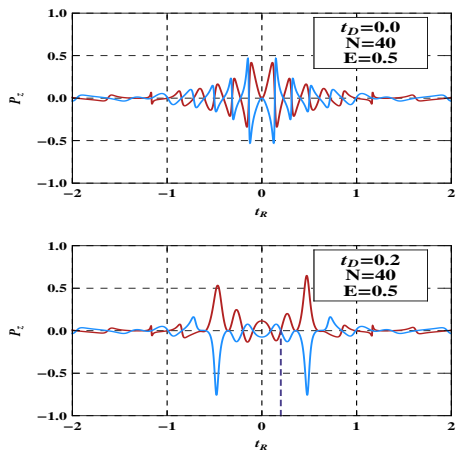}}\par}
\caption{(Color online). $P_z$ as a function of Rashba SO coupling. 
The lead-ring interface geometry and all the other parameters are 
same as Fig.~\ref{polx}.}
\label{polz}
\end{figure}
leads are no longer identical even when one SO coupling is set equal to
zero. This is solely due to the effect of quantum interference among the 
electronic waves passing through different arms of the mesoscopic ring.
All the other properties, for examples, the phase reversals of the 
polarization components $P_x$ and $P_z$ in two output leads and the 
vanishingly small amplitude of $P_y$ remain exactly same as discussed
earlier in the case of symmetric configuration.

The same quantities are also analyzed for this asymmetric configuration 
considering the ring with only Dresselhaus SO interaction, i.e., 
setting $t_R=0$. The results are given in Fig.~\ref{polasymeng2}.
Except getting different amplitudes of polarizing currents in two output
leads, all the other physical phenomena remain unchanged as discussed
in Fig.~\ref{poleng2}.

Before we end this section, in Figs.~\ref{polx}-\ref{polz} we describe the
variations of $P_x$, $P_y$ and $P_z$ as a function of Rashba SO coupling
considering two different values of $t_D$ for this asymmetric lead-ring
interface geometry to make the present communication a self contained
study. The amplitudes of the individual components of spin polarization
in two output terminals get changed, as expected, and their magnitudes 
can also be tuned by including other SO coupling. The physical picture
about the possibilities of determining SO fields by observing the 
vanishing spin polarization of out going currents in the limit $t_R=t_D$
remains also invariant for this lead-ring interface geometry, like the
symmetric configuration.

\section{Conclusion}

To conclude, in the present work we have described spin dependent
transport through a multi-terminal mesoscopic ring in presence of Rashba
and Dresselhaus SO interactions. Within a TB framework we have determined 
the polarization components $P_x$, $P_y$ and $P_z$ of outgoing currents using 
a general spin density matrix formalism. The sensitivity of these components
on the lead-ring interface geometry has also been analyzed in detail to
make the communication is self contained study. From our extensive numerical
work we have established that a two-terminal mesoscopic ring with only
SO coupling, in absence of any magnetic impurity or external magnetic field,
cannot induce spin polarization in the output lead. On the other hand,
a multi-terminal geometry, subjected to SO interaction, containing at least 
two out put leads can generate polarized spin currents from a completely 
unpolarized electron beam even in absence of any magnetic field or magnetic
like impurities. Finally, we have also provided a possible realization of
determining Dresselhaus SO coupling knowing the Rashba term in a single
sample by observing the vanishing spin polarization of the outgoing
currents, and hence facilitates a possible experimental measurement in
this line.

Finally we point out that, the presented results in this communication
are also valid even for non-zero temperatures ($\sim 300\,$K) as the 
broadening of the energy levels caused by side-attached leads is much
higher than the thermal broadening~\cite{datta1,datta2,sm1,sm2,sm3,sm4,sm5}.

\end{document}